\documentclass {entcs}
\usepackage [utf8] {inputenc}
\usepackage [T1] {fontenc}
\usepackage {url}

\begin {document}
\begin{frontmatter}
\title {Formalization of closure properties for context-free grammars}
\author{Marcus V. M. Ramos	
\thanksref{myemail}
}
\address{Centro de Informática\\UFPE\\Recife, Brazil} 
\thanks[myemail]{Email:\href{mailto:mvmr@cin.ufpe.br} {\texttt{\normalshape mvmr@cin.ufpe.br}}}	

\author{Ruy J. G. B. de Queiroz
\thanksref{coemail}
}
\address{Centro de Informática\\UFPE\\Recife, Brazil} 
\thanks[coemail]{Email:\href{mailto:ruy@cin.ufpe.br} {\texttt{\normalshape ruy@cin.ufpe.br}}}	
\begin {abstract}
Context-free language theory is a well-established area of mathematics, relevant to computer science foundations and technology. This paper presents the preliminary results of an ongoing formalization project using context-free grammars and the Coq proof assistant. The results obtained so far include the representation of context-free grammars, the description of algorithms for some operations on them (union, concatenation and closure) and the proof of related theorems (e.g. the correctness of these algorithms). A brief survey of related works is presented, as well as plans for further development.
\end {abstract}

\end{frontmatter}

\section {Introduction}
The fundamental mathematical activity of stating and proving theorems has been traditionally done by professionals that rely purely on their own personal efforts in order to accept or refuse any new proposal, after extensive checking. This style of work, which has been used for centuries by mathematicians all over the world, is now changing thanks to computer technology support. 

The so called ``proof assistants'' are software tools that are used in regular computers and offer a friendly and fully checked environment where one can describe mathematical objects and properties and then prove theorems about them. The main advantage of their use is to automate the validation of these demonstrations. When applied to program development, these tools are also helpful in checking the correctness of an existing program and in the construction of correct programs. In order to obtain these benefits, however, one must first be familiar with the underlying mathematical theory of each of these tools. 

Language theory is a well-established area in mathematics and computer science, which was extensively developed during the 1960s and 1970s. Automata theory came along and since the 1960s the two areas are generally considered as a single discipline. Fundamental to the study and development of computer languages, as well as computer languages processing technology, the theory also leads to important conclusions about the limits and properties of the computation process itself.

New and different uses of Coq and other proof assistants are announced frequently. These include, for example, the proof of the Four Color Theorem by Georges Gonthier and Benjamin Werner at Microsoft Research in 2005 \cite {gonthier-four} and also the demonstration of the Feit-Thompson Theorem by a group led by Georges Gonthier in 2012 \cite {gonthier-feit}. Also, there are important projects in the areas of mathematics \cite {arXiv-hales}, compiler certification \cite {leroy-2009} and digital security certification \cite {javacard}, among others \cite {coq-projects-lang} \cite {coq-projects-math}. 

The idea of formalizing context-free language theory in the Coq proof assistant is discussed in Section \ref {sec-formal-lang}. In particular, we present the goals that are being set, the strategy adopted and the results obtained so far. Finally, in Section \ref {sec-further} the plan for the rest of this research is presented and in Section \ref {sec-related} related work by various other researchers is considered.

\section {Formalization}
\label {sec-formal-lang}
The objective of this work is to formalize a substantial part of context-free language theory in the Coq proof assistant, making it possible to reason about it in a fully checked environment, with all the related advantages. Initially, however, the focus has been restricted to context-free grammars and associated results. Stack automata and their relation to context-free grammars shall be considered in the future.

More information on the Coq proof assistant, as well as on the syntax and semantics of the following definitions and statements, can be found in \cite {coq-site}, \cite {coq-2012} and \cite {bertot-2004}. 

The motivation for this work comes from (i) the large amount of formalization already existing for regular language theory; (ii) the apparent absence of a similar formalization effort for context-free language theory, at least in the Coq proof assistant and (iii) the high interest in context-free language theory formalization as a result of its practical importance in computer technology (e.g. correctness of language processing software). More information on related works is provided in Section \ref {sec-related}.

Context-free grammars have been represented in Coq very closely to the usual algebraic definition $G=(V,\Sigma,P,S)$, where $\Sigma$ is the set of terminal symbols (used in the construction of the sentences of the language generated by the grammar), $N=V-\Sigma$ is the set of non-terminal symbols (representing different sentence abstractions), $P$ is the set of rules and $S \in N$ is the start symbol (also called initial or root symbol). Rules have the form $\alpha \rightarrow \beta$, with $\alpha \in N$ and $\beta \in V^*$. The following record representation has been used:

\begin{verbatim}
Record cfg: Type:= {
non_terminal: Type;
terminal: Type;
start_symbol: non_terminal;
sf:= list (non_terminal + terminal);
rules: non_terminal -> sf -> Prop
}.
\end{verbatim}

The definition above states that \texttt {cfg} is a new type and contains four components. The first is \texttt {non\_terminal}, which represents the set of the non-terminal symbols of the grammar, the second is \texttt {terminal}, representing the set of terminal symbols, the third is \texttt {start\_symbol} and the fourth is \texttt {rules}, that represent the rules of the grammar. Rules are propositions (represented in Coq by \texttt {Prop}) that take as arguments a non-terminal symbol and a (possibly empty) list of non-terminal and terminal symbols (corresponding, respectively, to the left and right-hand side of a rule). \texttt {sf} (sentential form) is a list of terminal and non-terminal symbols.

Another fundamental concept used in this formalization is the idea of \emph {derivation}: a grammar \texttt{g} \emph {derives} a string \texttt {s2} from a string \texttt {s1} if there exists a series of rules in \texttt {g} that, when applied to \texttt {s1}, eventually result in \texttt {s2}. An inductive predicate definition of this concept in Coq uses two constructors:

\begin{verbatim}
Inductive derives (g: cfg): sf g -> sf g -> Prop :=
| derives_refl: forall s: sf g,
                derives g s s
| derives_step: forall s1 s2 s3: sf g,
                forall left: non_terminal g,
                forall right: sf g,
                derives g s1 (s2 ++ inl left :: s3)%list ->
                rules g left right ->
                derives g s1 (s2 ++ right ++ s3)%list.

\end{verbatim}

The constructors of this definition (\texttt {derives\_refl} and \texttt {derives\_step}) are the axioms of our theory. Constructor \texttt {derives\_refl} asserts that every sentential form \texttt {s} can be derived from \texttt {s} itself. Constructor \texttt {derives\_step} states that if a sentential form that contains the left-hand side of a rule is derived by a grammar, then the grammar derives the sentential form with the left-hand side substituted by the right-hand side of the same rule. This case corresponds the application of a rule in a direct derivation step.

Finally, a grammar \emph {generates} a string if this string can be derived from its root symbol:

\begin{verbatim}
Definition generates (g: cfg) (s: sf g): Prop:=
derives g [inl (start_symbol g)] s.
\end{verbatim} 

With these definitions, it has been possible to prove lemmas and also to implement functions that operate on grammars, all of which were useful when proving the main theorems. 

After context-free grammars and derivations were defined, the basic operations of concatenation, union and closure of context-free grammars were implemented in a rather straightforward way. These operations provide, as their name suggests, new context-free grammars that generate, respectively, the concatenation, the union and the closure of the language(s) generated by the input grammar(s). The code for these terms is presented below:

\noindent\emph {Union:}
\begin{verbatim}
Definition g_uni_t (g1 g2: cfg): Type:= 
(terminal g1 + terminal g2)%type.

Inductive g_uni_nt (g1 g2: cfg): Type :=
| Start_uni : g_uni_nt g1 g2
| Transf1_uni : non_terminal g1 -> g_uni_nt g1 g2
| Transf2_uni : non_terminal g2 -> g_uni_nt g1 g2.

Definition g_uni_sf_lift_left (g1 g2: cfg)
(c: non_terminal g1 + terminal g1): g_uni_nt g1 g2 + g_uni_t g1 g2:=
  match c with
  | inl nt => inl (Transf1_uni g1 g2 nt)
  | inr t  => inr (inl t)
  end.

Definition g_uni_sf_lift_right (g1 g2: cfg)
(c: non_terminal g2 + terminal g2): g_uni_nt g1 g2 + g_uni_t g1 g2:=
  match c with
  | inl nt => inl (Transf2_uni g1 g2 nt)
  | inr t  => inr (inr t)
  end.

Inductive g_uni_rules (g1 g2: cfg): g_uni_nt g1 g2 -> 
list (g_uni_nt g1 g2 + g_uni_t g1 g2) -> Prop :=
| Start1_uni: g_uni_rules g1 g2 (Start_uni g1 g2) 
              [inl (Transf1_uni g1 g2 (start_symbol g1))]
| Start2_uni: g_uni_rules g1 g2 (Start_uni g1 g2) 
              [inl (Transf2_uni g1 g2 (start_symbol g2))]
| Lift1_uni: forall nt s,
             rules g1 nt s ->
             g_uni_rules g1 g2 (Transf1_uni g1 g2 nt) 
             (map (g_uni_sf_lift_left g1 g2) s)
| Lift2_uni: forall nt s,
             rules g2 nt s ->
             g_uni_rules g1 g2 (Transf2_uni g1 g2 nt) 
             (map (g_uni_sf_lift_right g1 g2) s).

Definition g_uni (g1 g2: cfg): cfg := {|
non_terminal:= g_uni_nt g1 g2;
terminal:= g_uni_t g1 g2;
start_symbol:= Start_uni g1 g2;
rules:= g_uni_rules g1 g2
|}.
\end{verbatim}

The first definition above (\texttt {g\_uni\_t}) represents the type of the terminal symbols of the union grammar, created from the terminal symbols of the source grammars. Basically, it states that the terminals of both source grammars become terminals in the union grammar, by means of a disjoint union operation. For the non-terminal symbols, a more complex statement is required. First, the non-terminals of the source grammars are mapped to non-terminals of the union grammar. Second, there is the need to add a new and unique non-terminal symbol (\texttt {Start\_uni}), which will be the root of the union grammar. This is accomplished by the use of an inductive type definition (\texttt {g\_uni\_nt}), in contrast with the previous case, that used a simple non inductive definition.

The functions \texttt {g\_uni\_sf\_lift\_left} and \texttt {g\_uni\_sf\_lift\_right} simply map sentential forms from, respectively, the first or the second grammar in a pair, and produce sentential forms for the union grammar. This will be useful when defining the rules of the union grammar.

The rules of the union grammar are represented by the inductive definition \texttt {g\_uni\_rules}. Constructors \texttt {Start1\_uni} and \texttt {Start2\_uni} state that two new rules are added to the union grammar: respectively the rule that maps the new root to the root of the first grammar, and the rule that maps the new root to the root of the second grammar. Then, constructors \texttt {Lift1\_uni} and \texttt {Lift2\_uni} simply map rules in first (resp. second) grammar in rules of the union grammar.

Finally, \texttt {g\_uni} describes how to create a union grammar from two arbitrary source grammars. It uses the previous definitions to give values to each of the components of a new grammar definition.

Similar definitions were created to represent the concatenation of any two grammars and the closure of a grammar:

\noindent\emph {Concatenation:}
\begin{verbatim}
Definition g_cat_t (g1 g2: cfg): Type:= 
(terminal g1 + terminal g2)%type.

Inductive g_cat_nt (g1 g2: cfg): Type :=
| Start_cat : g_cat_nt g1 g2
| Transf1_cat : non_terminal g1 -> g_cat_nt g1 g2
| Transf2_cat : non_terminal g2 -> g_cat_nt g1 g2.

Definition g_cat_sf_lift_left (g1 g2: cfg)
(c: non_terminal g1 + terminal g1): g_cat_nt g1 g2 + g_cat_t g1 g2:=
  match c with
  | inl nt => inl (Transf1_cat g1 g2 nt)
  | inr t  => inr (inl t)
  end.

Definition g_cat_sf_lift_right (g1 g2: cfg)
(c: non_terminal g2 + terminal g2): g_cat_nt g1 g2 + g_cat_t g1 g2:=
  match c with
  | inl nt => inl (Transf2_cat g1 g2 nt)
  | inr t  => inr (inr t)
  end.

Inductive g_cat_rules (g1 g2: cfg): g_cat_nt g1 g2 -> 
list (g_cat_nt g1 g2 + g_cat_t g1 g2) -> Prop :=
| New_cat: g_cat_rules g1 g2 (Start_cat g1 g2) 
           ([inl (Transf1_cat g1 g2 (start_symbol g1))]++
           [inl (Transf2_cat g1 g2 (start_symbol g2))])%list
| Lift1_cat: forall nt s,
             rules g1 nt s ->
             g_cat_rules g1 g2 (Transf1_cat g1 g2 nt)
             (map (g_cat_sf_lift_left g1 g2) s)
| Lift2_cat: forall nt s,
             rules g2 nt s ->
             g_cat_rules g1 g2 (Transf2_cat g1 g2 nt)
             (map (g_cat_sf_lift_right g1 g2) s).

Definition g_cat (g1 g2: cfg): cfg := {|
non_terminal:= g_cat_nt g1 g2;
terminal:= g_cat_t g1 g2;
start_symbol:= Start_cat g1 g2;
rules:= g_cat_rules g1 g2
|}.
\end{verbatim}
		
\noindent\emph {Closure:}
\begin{verbatim}
Definition g_clo_t (g: cfg): Type:= 
(terminal g)%type.

Inductive g_clo_nt (g: cfg): Type :=
| Start_clo : g_clo_nt g
| Transf_clo : non_terminal g -> g_clo_nt g.

Definition g_clo_sf_lift (g: cfg)
(c: non_terminal g + terminal g): g_clo_nt g + g_clo_t g:=
  match c with
  | inl nt => inl (Transf_clo g nt)
  | inr t  => inr t
  end.

Inductive g_clo_rules (g: cfg): g_clo_nt g -> 
list (g_clo_nt g + g_clo_t g) -> Prop :=
| New1_clo: g_clo_rules g (Start_clo g) 
            ([inl (Start_clo g)]++
             [inl (Transf_clo g (start_symbol g))])
| New2_clo: g_clo_rules g (Start_clo g) []
| Lift_clo: forall nt s,
            rules g nt s ->
            g_clo_rules g (Transf_clo g nt) (map (g_clo_sf_lift g) s).

Definition g_clo (g: cfg): cfg := {|
non_terminal:= g_clo_nt g;
terminal:= g_clo_t g;
start_symbol:= Start_clo g;
rules:= g_clo_rules g
|}.
\end{verbatim}

Although simple in their structure, it must be proved that the definitions \texttt {g\_uni}, \texttt {g\_cat} and \texttt {g\_clo} always produce the correct result. In other words, the algorithms embedded in these definitions must be ``certified''. The process of doing such a certification is called ``program verification'', and is one of the main goals of formalization. In order to accomplish this, we must first state theorems, using first-order logic, that capture the expected semantics of these definitions. Finally, we have to derive proofs of the correctness of these theorems. 

This can be done with a pair of theorems for each definition/algorithm: the first relates the output to the inputs, and the other one does the inverse, providing assumptions about the inputs once an output is generated. This is necessary in order to guarantee that the algorithm does only what one would expect, and no more.

\noindent\emph {Concatenation, direct operation:}
\begin{verbatim}
Theorem g_cat_correct (g1 g2: cfg)(s1: sf g1)(s2: sf g2):
generates g1 s1 /\ generates g2 s2 ->
generates (g_cat g1 g2)
          ((map (g_cat_sf_lift_left g1 g2) s1)++
          (map (g_cat_sf_lift_right g1 g2) s2))%list.
\end{verbatim}

The above theorem, for example, states that if context-free grammars \texttt{g1} and \texttt{g2} generate, respectively, strings \texttt{s1} and \texttt{s2}, then the concatenation of these two grammars, according to the proposed algorithm, generates the concatenation of string \texttt{s1} to string \texttt{s2}. As mentioned before, the above theorem alone does not guarantee that \texttt{g\_cat} will not produce outputs other than the concatenation of its input strings. This idea is captured by the following complementary theorem:

\noindent\emph {Concatenation, inverse operation:}
\begin{verbatim}
Theorem g_cat_correct_inv (g1 g2: cfg)(s: sf (g_cat g1 g2)):
generates (g_cat g1 g2) s ->
exists s1: sf g1, 
exists s2: sf g2,
s =(map (g_cat_sf_lift_left g1 g2) s1)++
   (map (g_cat_sf_lift_right g1 g2) s2) /\
generates g1 s1 /\
generates g2 s2.
\end{verbatim}

The idea here is to express that, if a string is generated by \texttt{g\_cat}, then it must only result from the concatenation of strings generated by the grammars merged by the algorithm. Together, these two theorems represent the semantics of the context-free grammar concatenation operation presented. The same ideas have been applied to the statement and proof of the following theorems, relative to the union and closure operations:

\noindent\emph {Union, direct operation:}
\begin{verbatim}
Theorem g_uni_correct (g1 g2: cfg)(s1: sf g1)(s2: sf g2):
generates g1 s1 \/ generates g2 s2 ->
generates (g_uni g1 g2) (map (g_uni_sf_lift_left g1 g2) s1) \/ 
generates (g_uni g1 g2) (map (g_uni_sf_lift_right g1 g2) s2).
\end{verbatim}

\noindent\emph {Union, inverse operation:}
\begin{verbatim}
Theorem g_uni_correct_inv (g1 g2: cfg)(s: sf (g_uni g1 g2)):
generates (g_uni g1 g2) s ->
(s=[inl (start_symbol (g_uni g1 g2))]) \/
(exists s1: sf g1,
(s=(map (g_uni_sf_lift_left g1 g2) s1) /\ generates g1 s1)) \/
(exists s2: sf g2,
(s=(map (g_uni_sf_lift_right g1 g2) s2) /\ generates g2 s2)).
\end{verbatim}

\noindent\emph {Closure, direct operation:}
\begin{verbatim}
Theorem g_clo_correct (g: cfg)(s: sf g)(s': sf (g_clo g)):
generates (g_clo g) nil /\
(generates (g_clo g) s' /\ generates g s -> 
generates (g_clo g) (s'++ (map (g_clo_sf_lift g)) s)).
\end{verbatim}

\noindent\emph {Closure, inverse operation:}
\begin{verbatim}
Theorem g_clo_correct_inv (g: cfg)(s: sf (g_clo g)):
generates (g_clo g) s -> 
(s=[]) \/
(s=[inl (start_symbol (g_clo g))]) \/
(exists s': sf (g_clo g), 
 exists s'': sf g,
 generates (g_clo g) s' /\ generates g s'' /\ 
 s=s'++map (g_clo_sf_lift g) s'').
\end{verbatim}

The proofs of all the six main theorems have been completed (\texttt {g\_uni\_correct} and \texttt {g\_uni\_correct\_inv} for union, \texttt {g\_cat\_correct} and \texttt {g\_cat\_correct\_inv} for concatenation and \texttt {g\_clo\_correct} and \texttt {g\_clo\_correct\_inv} for closure). As an interesting side result, some useful and generic lemmas have also been proved during this process. Among these, for example, one that asserts the context-free characteristic of these derivations:

\begin{verbatim}
Theorem derives_context_free_add:
forall g:cfg,
forall s1 s2 s s': sf g,
derives g s1 s2 -> derives g (s++s1++s') (s++s2++s').
\end{verbatim}

\noindent
and one that states the transitivity of the \texttt {derives} relation:

\begin{verbatim}
Theorem derives_trans:
forall g: cfg,
forall s1 s2 s3: sf g,
derives g s1 s2 ->
derives g s2 s3 ->
derives g s1 s3.
\end{verbatim}

All the definitions and proof scripts were written in plain Coq using CoqIDE (a graphical interface for Windows), and are available for download at: \\
\url {http://www.univasf.edu.br/~marcus.ramos/coq/cfg-closure.v}. \\
Basically, the proof scripts use induction on the predicate \texttt {derives} and also direct list manipulation. The libraries used were \texttt{Ascii}, \texttt{String} and \texttt{List}.

\section {Further Work}
\label {sec-further}

The proper specification of inductive predicates and related definitions leads naturally to simple functions that promote the necessary transformations on the objects described, and also to readable statements of lemmas and theorems.

The current work concentrated on the formalization of context-free grammars, derivations and closure operations, as well as on the certification of the correctness of these operations. The plan now is to use the same definitions to achieve the following goals:

\begin {enumerate}
\item Describe algorithms for the simplification of context-free grammars (namely elimination of inaccessible and useless symbols, unit and empty productions) and prove their correctness;
\item Similarly for the construction of Chomsky and Greibach normal forms for context-free grammars;
\item Prove some decidable questions on context-free languages, especially those whose proofs rely on context-free grammars;
\item Finally, obtain a formal proof of the Pumping Lemma for context-free languages, and to use it to prove the existence of non context-free languages. 
\end {enumerate}

All the theory and results related to stack automata, and also to the relation of stack automata to context-free grammars, shall be left for future work.

\section {Related Work}
\label {sec-related}
Language and automata theory has been subject of formalization since the mid-1980s, when Kreitz used the Nuprl proof assistant to prove results about  deterministic finite automata and the pumping lemma for regular languages \cite {kreitz-1986}. Since then, the theory of regular languages has been formalized partially by different researchers using different proof assistants (see \cite {constable-1997}, \cite {kaloper-1996}, \cite {filliatre-1997a}, \cite {briais-2008}, \cite {miyamoto}, \cite {moreira-2009}, \cite {almeida-2010a}, \cite {almeida-2010b}, \cite {moreira-2012} \cite {braibant-2010}, \cite {braibant-2012}, \cite {asperti-2012b}, \cite {coquand-2011}, \cite {krauss-2012} and \cite {wu-2011}). The most recent and complete formalization, however, is the work by Jan-Oliver Kaiser \cite {doczkal-2013}, which used Coq and the SSReflect extension to prove the main results of regular language theory.

Context-free language theory has not been formalized the same extent so far. The more relevant works are the ones published by Jourdan, Pottier and Leroy (using Coq, \cite {jourdan-2012}) and Ridge (HOL4, \cite {ridge-2011}), on parser generation and validation, and Norrish and Barthwal (HOL4, \cite {barthwal-norrish-2010a}, \cite {barthwal-norrish-2010b}, \cite {barthwal-norrish-2013}), on theory formalization.

When it comes to computability theory and Turing machines related classes of languages, formalization bas been approached by Asperti and Ricciotti (Matita, \cite {asperti-2012b}), Xu, Zhang and Urban (Isabelle/HOL, \cite {xu-2013}) and Norrish (HOL4, \cite {norrish-2011}).

\section {Conclusions}

The present paper reports an ongoing research effort towards formalizing the classical context-free language theory, initially based only on context-free grammars, in the Coq proof assistant. All important objects have already been described and basic closure operations on grammars have already been implemented. Proofs of the correctness of the concatenation, union and closure operations (for both direct and inverse ways) were constructed.

When the work is complete, it should be useful for a few different purposes. Among them, to offer a complete and mathematically precise description of the behavior of the objects of context-free language theory. Second, to offer fully checked and mechanized demonstrations of its main results. Third, to allow for the certified and efficient implementation of its relevant algorithms in a proper programming language. Fourth, to permit the experimentation in an educational environment in the form of a tool set, in a laboratory where further practical observations and developments can be done, for the benefit of students, teachers, professionals and researchers.

\bibliographystyle {entcs}
\bibliography {article}
\end {document}